# Proposal of a Contact Detection System using Microphones for a Chambara-based Augmented Sports

Yusaku MAEDA[1], Sho SAKURAI[1], Koichi HIROTA[1], and Takuya NOJIMA[1]

[1] *The University of Electro-Communications, Tokyo, Japan*

({maeda_yusaku, sho, hirota}@vogue.is.uec.ac.jp, tnojima@nojilab.org)

**Abstract ---** This study presents a novel contact detection system for "Parablade," a chambara-based, sword-play augmented sport. Augmented sports combine physical activities with virtual parameters (VPs) to create a balanced and equitable gaming experience, irrespective of players' physical capabilities. The proposed Parablade Microphone Unit (PMU) employs multiple microphones and machine learning algorithms to detect and classify hit events through sound recognition. This system aims to ensure real-time updates of VPs, thereby enhancing the gameplay experience. Experimental results indicate that the PMU can accurately recognize the occurrence and location of hit events with a high accuracy rate of 93.33%, with the assistance of 10kHz additional sound generated from the sword.

**Keywords:** Augmented Sports, Chambara, Parablade, Sword Play, Hit Detection

## 1 INTRODUCTION

Augmented Sports represent a form of sports that combines existing physical sports with virtual parameters (VPs). VPs are commonly used in digital games, such as Hit Point (HP), Attack Point (ATK), and Defense Point (DEF). In Augmented Sports, such VPs are used to mitigate the negative effects of differences in physical abilities among participants[1]. VPs can enhance the player's ability in the Augmented Sports they play. For example, if a player is physically weak, increasing HP and ATK will result in making the player more powerful in the matching. Thus, well-designed VPs have enough potential to bridge the gap between each player's physical capabilities, thereby achieving equitable gameplay and enhancing enjoyment. To make VPs work effectively when playing Augmented Sports, a system that could detect various events during the sport is necessary to update VPs' status according to the ongoing matching. In this report, we introduce a microphone-based system that is capable of detecting events related to contact between the Augmented Sports player's body and objects that the player uses for play.

In a previous study, we developed Augmented Dodgeball[1], which is a physical dodgeball modified based on the Augmented Sports concept. The Augmented Dodgeball introduced three rolls with different VPs consisting of HP/ATK/DEF. A roll with higher HP and DEF resulted in lower damage taken and higher endurance, while another roll with higher ATK is capable of inflicting greater damage against opponents. Selecting a role that suits the player's characteristics helps to compensate for the player's physical ability and increase enjoyment. VPs have a significant impact on the winning probability in the game. For example, a team with a player who has an extensive HP will not lose, regardless of their original physical performance. Thus, a sophisticated VP design is essential to maintain and improve game enjoyment. At the same time, a system that can detect certain events and update VPs in real time is necessary to play augmented sports. For example, in the case of augmented dodgeball, if a physical ball hits a player, that event must be detected, and the player's HP must be decreased immediately according to the player's DEF and the thrower's ATK value.

Unfortunately, Augmented dodgeball remains the only example that shows the concept of Augmented Sports. Developing various Augmented Sports is necessary to validate the generality of the concept. Therefore, in this study, we aim to develop a new type of Augmented Sport named "Parablade," a chambara-based Augmented Sport, which is a sword-matching sport with VPs. As mentioned earlier, an appropriate event detection system is essential for the implementation of Augmented Sports, and this requirement is equally pertinent to the Parablade. Parablade is a sword sport that requires two participants. If one player hits a certain part of the opponent's body, the information on the occurrence of the hit event and the body part that was hit is necessary to update VPs. Similar to actual, well-played chambara, assumed damage is different according to the location of the body being hit by the sword. The player receives more extensive damage when being hit on the stomach compared to the case of being hit on the arm. To achieve this, we developed a microphone-based contact detection system named PMU (Parablade Microphone Unit). PMU consists of two microphones with a transmitter that is used to detect nearby hit-specific sounds. In this article, we describe the detailed structure of the developed PMU and its capability of detecting hitting events.

## 2 RELATED WORKS

The demands for more accurate judgment have led to the development of various refereeing systems using electronic equipment. For example, Hawk-Eye[2] is an image-based system used in ball games such as tennis to determine whether a ball has crossed a line. Unfortunately, image-based systems have potential difficulty in detecting contact event that occurs in the image sensor's blind spot.

One of the well-known contact detection systems used in the sports field is for fencing. The judgment system for fencing is designed to detect contact events between a player and a sword by establishing an electrical circuit through the metal sword[3]. Sword is a well-used device in sports, and various types of contact detection systems have been developed. SMART FENCING[4] uses a sensor and transmitter integrated sword and conductive vest to detect stabbing with the tip of the sword wirelessly. SASSEN[5] detects contact by using a pressure sensor integrated sword. Apart from the sports field, Takahashi et al. [6] proposed a contact detection technology between a ball-shaped device and the human body using personal area network (PAN) technology. Suzuki et al. [7] also proposed a device called EnhancedTouch, which can detect contact between humans. Although these systems can detect contact between humans to objects and/or other humans, they cannot determine the location of contact on the body. When playing existing chambara, the level of damage varies based on the part where the opponent's sword hits. For example, if the sword hits the opponent's arm, the opponent can still continue to play, but if the sword hits the opponent's body, the opponent receives more damage.

## 3 DEVELOPMENT OF A CONTACT DETECTION SYSTEM

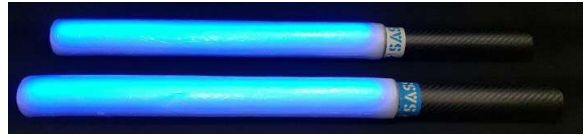

Fig.1 SASSEN sword

In this chapter, we describe the details of the contact detection system. This system is designed for Parablade, chambra-based augmented sports. For ease of play, the system should be easily equipable and can reliably detect the contact wirelessly in real time. Then, we propose a microphone-based contact detection system named PMU (Parablade Microphone Unit, Fig.2 ). Based on Machine Learning Technology, PMU detects hit-specific sounds around the device. Then, the player wears multiple PMUs on the body to detect the hitting sound generated on each part of the body. Unfortunately, the hitting sound will vary according to the clothes players wear and the swords they use. In this system, no specific clothing was specified, and we decided to develop the system under the rule of using a SASSEN sword (Fig.1 ) as a sword.

### 3.1 PMU (Parablade Microphone Unit)

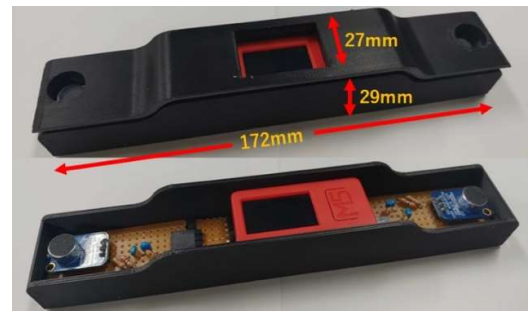

Fig.2 PMU (Parablade Microphone Unit)

The PMU is a microphone unit consisting of two microphones (Electret Microphone Amplifier - MAX4466 with Adjustable Gain ) and one microcontroller (M5Stack Technology Co., Ltd, M5StickCPlus). The microphones send the hitting sound signal to the microcontroller through a low-pass filter. The power spectrum of the recorded hitting sound when using the SASSEN sword shows that the distinctive feature between the human and the sword occurs between 50 Hz and 200 Hz. Consequently, we set the cut-off frequency at 252 Hz. Sound recognition on a microcontroller is implemented using TensorFlow Lite for Microcontrollers(tflite-micro)[8], a machine learning platform for microcontrollers, and is envisioned to recognize input from a connected microphone sensor. Then, by recording thousands of hit sounds and applying machine learning technology, the PMU should be able to be used as a hit-detection system.

### 3.2 Recording PMU for feasibility study

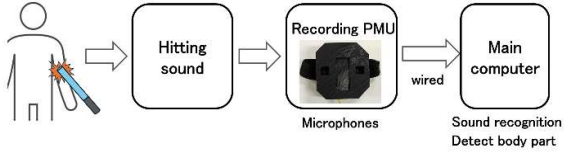

Fig.3 The workflow of the hitting detection using the recording PMU

Before integrating the full-spec PMU as described in 3.1 , we develop a recording PMU, a simplified version of PMU, for a feasibility study. This recording PMU is used throughout the document. In this version, hitting sound recognition is performed on a conventional notebook computer (Intel Core i7-10750H, GeForce RTX 3060 Laptop, RAM 16GB) for ease of development. The recording PMU functions solely as a microphone unit. The Fig.3 shows the workflow when using the recording PMU. When the SASSEN sword hits the human body, this leads to generating hitting sound intrinsic to the physical interaction. The recording PMU transmits this sound directly to the PC, bypassing the microcontroller. The main computer is used for identifying the occurrence of the hitting event, while the training of the recognition model with the hitting sound was performed on a desktop computer (GeForce RTX 3080, RAM 16GB).

## 4 HIT SOUND LEARNING AND EVALUATION FOR PMU

### 4.1 Learning for Hitting Sound

The PMU's hitting sound recognition model is based on tflite-micro's sample program of "speech commands." It consists of a two-dimensional convolutional neural network (2D-CNN) with DepthwiseConv2D (8 filters, 10x1 kernel, 2x2 stride, depth 8, bias, and ReLU activation), FullyConnected Layer (4 units, weights 4x4000, bias), and Softmax Layer (4 units, beta 1). As a preprocessing step, an FFT was performed on one second of raw sound data using a window function with a window size of 30ms and a window interval of 20ms to generate a spectrogram with a width of 43 and a height of 49. The sound data used for training was divided into training data, validation data, and test data at a ratio of 8:1:1, and the learning rate was set at 0.001 for the first 12,000 training loops and 0.0001 for the last 3,000 training loops, for the total of 15,000 training loops.

### 4.2 Data Set Collection

The training dataset was obtained by recording the sound hitting the human body with the SASSEN sword in various conditions. In this experiment, the recording PMU was embedded in the participant's left forearm and hit the participant's body with the SASSEN sword by himself. Each sound file was saved as a 1-second WAV file at a sampling frequency of 44.1 kHz, 16bit, in stereo format. In this study, we recorded the sound of hitting the participant's hand, forearm and upper arm. In addition, hitting another SASSEN sword, as well as silence and other sounds such as environmental noise (footsteps and conversations etc.) were recorded. Finally, a total of six patterns of sound were recorded via PMU on the forearm. The number of participants who were being hit was one. To arrange variations of recording data, the recording was conducted in three different locations (in the laboratory/outdoors/judo hall) and three different clothing (plain/t-shirt/hoodie). For the recording condition, the laboratory was a quiet environment with about five people operating PCs, the outdoor was noisy environment due to wind and passersby on campus, and the judo hall was an environment where sound tends to reverberate. In addition, the angle of the sword and strength were randomly changed when hitting to improve the generality of recording data. The data set was created by recording 1200 files for each of the five classes except silence, and 540 files for silence.

### 4.3 Classification accuracy after training

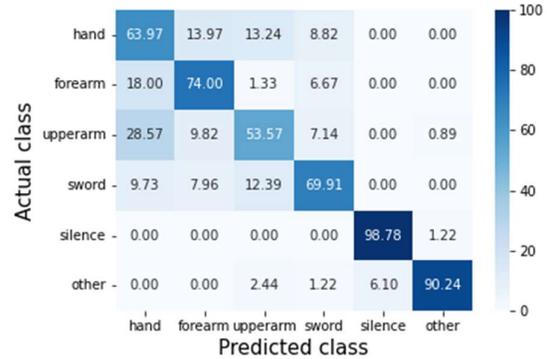

Fig.4 Sound Classification Accuracy Confusion Matrix

Fig.4 shows the confusion matrix of the training results. Among hitting conditions, i.e., hitting the hand or forearm or the upper arm, the maximum classification accuracy in training was 74.00%. Furthermore, by combining the result of three classes of hand, forearm, and upper arm as a new class named "body", the classification accuracy is 92.21%. "Silence" and "Other" were correctly classified with high accuracy at 98.78% and 90.24%, respectively. This indicates that the system has reasonable potential to detect the occurrence of hitting events. At the same time, an improvement is still necessary for classifying the location where the event occurs.

### 4.4 Adding sound information

To improve the accuracy of classification, an additional clue to identify the hitting event was investigated. In this trial, a speaker was placed at the tip of the SASSEN sword to continuously generate 10 kHz sinusoidal sound at a constant amplitude. The frequency of 10 kHz is within the audible range for humans; thus, it may burden players if it is used in actual matches. However, since the system is still proof of concept, the frequency was chosen because of ease of development. The recording was made again with this device, and five patterns were studied: hitting the hand/forearm/upper arm, the silent state, and others. In this experiment, the sword condition, hitting sword each other, was removed. This is because SASSEN swords are equipped with pressure sensors capable of detecting impacts as an inherent feature. By utilizing this function, the occurrence of hitting sword could be detected without the need for a PMU.

### 4.5 Classification accuracy by re-training

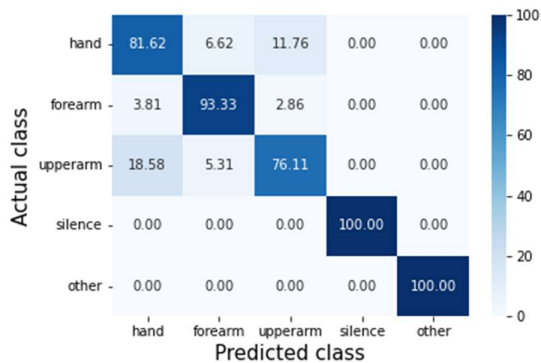

Fig.5 Sound Classification Accuracy with 10 kHz sound Confusion Matrix

Fig.5 shows the confusion matrix of the classification using the updated system. Consistent with the previous experiment, the PMU was installed on the forearm. The highest accuracy 93.33% was observed at the condition of the forearm, where the PMU is installed. Moreover, the accuracy tended to decrease as the hitting position moved further from the forearm, with accuracies of 81.62% and 76.11% for the hand and upper limb, respectively. These results are higher than those of the previous experiment. The overall average classification accuracy across the five conditions was 88.1%. This suggests that integrating a 10 kHz sound generator contributes to improving classification accuracy.

## 5 CONCLUSION AND FUTURE WORK

In this research, we propose a contact detection system for "Parablade," a physical sword-matching game that extends chambara as an Augmented Sport. The system named PMU utilizes multiple microphones to recognize the sword contact events through sound-based analysis. The experimental results showed that PMU can correctly detect hitting events with an accuracy of 93.33% with assistive audio cues. However, the system's performance may be influenced by factors such as the player's body posture, noise, and other environmental conditions. During trials in noisy settings, the recognition accuracy was occasionally compromised. As future work, we plan to assess the system's validity in actual gameplay settings with varying noise conditions to enhance its robustness.

ACKNOWLEDGEMENT

We gratefully acknowledge the invaluable support provided by the JAPAN SASSEN Association in the execution of this research.